\documentclass{raa}
\usepackage{graphicx,times}
\usepackage{natbib}
\usepackage{amsmath}
\usepackage{float,color}
\usepackage{savefnmark,booktabs}
\usepackage{threeparttable,supertabular}
\usepackage{setspace}

\begin{document}

\title{Intensity Distribution Function and Statistical Properties of Fast Radio Bursts 
}

\volnopage{ {\bf XXXX} Vol.\ {\bf X} No. {\bf XX}, 000--000}
\setcounter{page}{1}

\author{Long-Biao Li\inst{1,2}, Yong-Feng Huang\inst{1,2}$^{*}$, Zhi-Bin Zhang\inst{3}, Di Li\inst{4,5}, Bing Li\inst{1,2,6}  }

\institute{School of Astronomy and Space Science, Nanjing University, Nanjing 210046, China {\it hyf@nju.edu.cn}\\
    \and
          Key Laboratory of Modern Astronomy and Astrophysics (Nanjing University), Ministry of Education, Nanjing 210046, China \\
	\and
          Department of Physics, College of Sciences, Guizhou University, Guiyang 550025, China \\
    \and
          National Astronomical Observatories, Chinese Academy of Sciences, Beijing 100012, China \\
    \and
          Key Laboratory for Radio Astronomy, Chinese Academy of Sciences, China \\
    \and  Institute of High Energy Physics, Chinese Academy of Sciences, Beijing 100049, China \\
\vs \no
   {\small Received XXXX ; Accepted XXXX}
}

\abstract{
  Fast Radio Bursts (FRBs) are intense radio flashes from the sky that are
  characterized by millisecond durations and Jansky-level flux densities.
  We carried out a statistical analysis on FRBs discovered.
  Their mean dispersion measure, after subtracting the contribution from the
  interstellar medium of our Galaxy, is found to be $\sim 660\,\rm pc\,cm^{-3}$,
  supporting their being from cosmological origin.
  Their energy released in radio band spans about two orders of magnitude,
  with a mean value of $\sim 10^{39}$ ergs.
  More interestingly, although the FRB study is still in a very early phase,
  the published collection of FRBs enables us to derive a useful intensity distribution function.
  For the 16 non-repeating FRBs detected by Parkes telescope and the Green Bank Telescope,
  the intensity distribution can be described as
  $dN/dF_{\rm obs} = (4.1 \pm 1.3) \times 10^3 \, F_{\rm obs}^{-1.1\pm0.2} \; \rm sky^{-1}\,day^{-1}$,
  where $F_{\rm obs}$ is the observed radio fluence in units of Jy~ms.
  Here the power-law index is significantly flatter than the expected value of 2.5
  for standard candles distributed homogeneously in a flat Euclidean space.
  Based on this intensity distribution function, the Five-hundred-meter Aperture Spherical radio Telescope (FAST)
  will be able to detect about 5 FRBs for every 1000 hours of observation time.
}

   \authorrunning{Li, Huang, Zhang, Li \& Li}            
   \titlerunning{Intensity Distribution Function and Statistical Properties of FRBs}  
   \maketitle

\section{Introduction}            
\label{sect:intro}

Fast Radio Bursts (FRBs) are intense radio flashes that seem to occur randomly on the sky.
They are characterized by their high brightness ($\geq\,1\,\rm Jy$), but with very short durations ($\sim ms$).
Until March 2016, 16 non-repeating bursts and one repeat source have been discovered
as unexpected outcome of reprocessing pulsar and radio transient surveys
(Lorimer et al. 2007; Keane et al. 2012; Thornton et al. 2013; Burke-Spolaor \& Bannister 2014;
Spitler et al. 2014; Champion et al. 2016; Masui et al. 2015; Petroff et al. 2015a; Ravi, Shannon \& Jameson 2015;
Keane et al. 2016; Scholz et al. 2016; Spitler et al. 2016).
Except for the possible counterpart and host galaxy of FRB 150418 identified in Keane et al. 2016
(but see Vedantham et al. 2016 and Williams \& Berger 2016 for different opinions),
most previous efforts trying to search for the counterparts of other FRBs have led to a negative
result (e.g. Petroff et al. 2015a). Very recently, \citet{Spitler2016} and \citet{Scholz2016}
discovered sixteen repeating bursts from the direction of FRB 121102, providing valuable
clues to the nature of these enigma events (Dai et al. 2016; Gu et al. 2016).
Although FRB 140514 has been detected in almost the same direction as that of FRB 110220, it is
considered to be a separate event because of its different dispersion measure (DM) (Petroff et al. 2015b).

The arrival time of a FRB at different wavelength is characterized by a frequency-dependent delay of $\Delta t \propto \nu^{-2}$,
and the pulse width is found to scale as $W_{\rm obs} \propto \nu^{-4}$ (Lorimer et al. 2007; Thornton et al. 2013).
Both characteristics are consistent with the expectations for radio pulses propagating
through a cold, ionized plasma. These facts strengthen the view that FRBs are of astrophysical origin.
The dispersion measure, defined as the line-of-sight integral of the free electron number
density, is a useful indication of distance. An outstanding feature of FRBs is that their
dispersion measures are very large and exceed the contribution from
the electrons in our Galaxy by a factor of 10 --- 20 in most cases.
\citet{Lorimer2007} and \citet{Thornton2013} suggested that the large DM is dominated by the
contribution from the ionized intergalactic medium (IGM).
FRBs thus seem to occur at cosmological distances. With their large DMs, FRBs may be
a powerful probe for studying the IGM and the spatial distribution of free electrons.

The millisecond duration of FRBs suggests that their sources should be compact,
and the high radio brightness requires a coherent emission mechanism (Katz 2014a; Luan \& Goldreich 2014).
Since FRBs' redshifts are estimated to be in the range of $z \sim 0.5$ --- $1.3$
(Thornton et al. 2013; Champion et al. 2016),
the emitted energy at radio wavelengths can be as high as $\sim 10^{39}$ --- $10^{40}$ ergs.
Although the physical nature of FRBs is still unclear,
some possible mechanisms have been proposed, such as double neutron star mergers (Totani 2013),
interaction of planetary companions with the magnetic fields of pulsars (Mottez \& Zarka 2014),
collapses of hypermassive neutron stars into black holes (Falcke \& Rezzolla 2014;
Ravi \& Lasky 2014; Zhang 2014),
magnetar giant flares (Kulkarni et al. 2014; Lyubarsky 2014; Pen \& Connor 2015),
supergiant pulses from pulsars (Cordes \& Wasserman 2016),
collisions of asteroids with neutron stars (Geng \& Huang 2015; Dai et al. 2016)
or the inspiral of double neutron stars (Wang et al. 2016).
Keane et al. (2016) suggested that there may actually be more than one class of FRB progenitors.

New FRB detections are being made and much more are expected in the near future.
\citet{Thornton2013} have argued that if FRBs happen in the sky isotropically, their actual event rate
could be as high as $\sim 10^4\,\rm sky^{-1} day^{-1}$. \citet{Hassall2013}
discussed the prospects of detecting FRBs with the next-generation radio telescopes
and suggested that the Square Kilometre Array (SKA) could detect about one FRB per hour.
Based on the redshifts estimated from the measured DMs, \citet{Bera2016}
studied the FRB population and predicted that the upcoming Ooty Wide Field
Array\footnote{http://rac.ncra.tifr.res.in/} can detect FRBs at a rate of
$\sim 0.01$ --- $10^3$ per day, depending on the power-law index
of the assumed distribution function, which could vary from -5 to 5.
Note that their predicted detection rate is in a very wide range,
which mainly stems from the uncertainty of the FRB luminosity function and their spectral index.

The luminosities depends strongly on the measured redshifts.
However, the redshifts of FRBs are not directly measured, but are derived from their DMs.
The reliability of these redshifts still needs to be clarified (Katz 2014b; Luan \& Goldreich, 2014; Pen \& Connor 2015).
In this study, we examine the statistical properties of published FRBs,
and use the directly measured fluences\footnote{Note that the usage of the word "fluence"
here is different from its common definition and dimension.
We follow earlier FRB papers in this study.}
of FRBs to derive an intensity distribution function.
Our distribution function is independent of the redshift measurements.
We then use the intensity distribution function to estimate the detection rate of FRBs by
the Chinese Five-hundred-meter Aperture Spherical radio Telescope (FAST),
the opening ceremony of which is slated for the 25th of September, 2016.
Our article is organized as follows.
In Section 2, we describe the sample of 16 non-repeating FRBs
and present the statistical analyses of their parameters.
In Section 3, we derive the intensity distribution function of FRBs.
In Section 4, the observational prospects of FRBs with FAST are addressed.
Our conclusions and discussion are presented in Section 5.

\section{Samples and Statistical Analyses}

We extract the key parameters of 16 non-repeating FRBs detected by Parkes and GBT
from the FRB Catalogue of \citet{Petroff2016} \footnote{http://astronomy.swin.edu.au/pulsar/frbcat/}.
The data are listed in Table \ref{tab1}. At the direction of FRB 121102, additional 16 repeating bursts
have been detected (Spitler et al. 2016; Scholz et al. 2016), indicating that all these events may be quite different
from other non-repeating FRBs in nature. So we treat these 17 repeating events separately in our following
study.

Column 1 of Table \ref{tab1} provides the FRB names.
The observed width or duration of the corresponding radio pulse ($W_{\rm obs}$)
is presented in Column 2.
Column 3 is the observed peak flux density ($S_{\rm peak}$) of each FRB.
Column 4 tabulates the observed fluences ($F_{\rm obs}$) in units of Jy ms,
which is calculated as $F_{\rm obs} = S_{\rm peak} \times  W_{\rm obs}$.
Columns 5, 6, and 7 present the observed DMs of FRBs,
the DM contributions from the Galaxy ($ DM_{\rm Galaxy}$), and the DM
excesses ($DM_{\rm Excess}$), respectively. The DM excess is defined
as $DM_{\rm Excess} = DM - DM_{\rm Galaxy}$.
The estimated redshift ($z$) is presented in Column 8, assuming that the density of electrons is a constant for the IGM.
The corresponding luminosity distances ($D_{\rm L}$) and the emitted energies ($E$)
are presented in Columns 9 and 10, respectively.
Note that there is no reliable estimate on the uncertainties of $DM_{\rm Galaxy}$,
therefore, it is not included. Then the uncertainties of $DM_{\rm Excess}$, $z$, $D_{L}$ and $E$ are also not available.

\begin{table*}
  \caption{Key parameters of the 16 non-repeating FRBs. Observational data are mainly taken from
  http://astronomy.swin.edu.au/pulsar/frbcat/ (Petroff et al. 2016). \label{tab1}}
  \centering
  \scalebox{0.87}{
  \begin{threeparttable}
  \begin{spacing}{1.55}
  \begin{tabular}{@{}lcccccccccc@{}}
  \hline
  \hline
    FRB  & $W_{\rm obs}$  & $S_{\rm peak}$ & $F_{\rm obs}$ &   $DM$\tnote{a} &   $ DM_{\rm Galaxy}$\tnote{a}
                                         & $ DM_{\rm Excess}$\tnote{b} &  $z$\tnote{c}  & $D_{\rm L}$\tnote{c}     & $E$\tnote{c}  \\
     &       (ms)     &      (Jy)      &  (Jy ms)  & (${\rm pc\,cm^{-3}}$) & (${\rm pc\,cm^{-3}}$)
                                         & (${\rm pc\,cm^{-3}}$) &   &   (Gpc)    & ($10^{39}$ ergs)         \\
  (1)  & (2) & (3) & (4) &  (5)  &  (6)  &  (7) & (8)  &  (9)  &  (10)     \\
  \hline
  010125  &  10.60$^{+2.80}_{-2.50}$    &  0.54$^{+0.11}_{-0.07}$   &  5.72$^{+2.99}_{-1.92}$   &  $790.3\pm0.3$    &  110   &  680.3   &  0.57  &  3.35  &  2.77     \\
  010621  &  8.00 $^{+4.00}_{-2.25}$    &  0.53$^{+0.26}_{-0.09}$   &  4.24$^{+5.24}_{-1.71}$   &  $748\pm3$        &  523   &  223     &  0.19  &  0.93  &  0.12     \\
  010724  &  20.00$^{+0.00}_{-0.00}$    &  1.57$^{+0.00}_{-0.00}$   &  31.48                    &  $375\pm3$        &  44.58 &  330.42  &  0.28  &  1.45  &  2.31     \\
  090625  &  1.92 $^{+0.83}_{-0.77}$    &  1.14$^{+0.42}_{-0.21}$   &  2.19$^{+2.10}_{-1.12}$   &  $899.55\pm0.1$   &  31.69 &  867.86  &  0.72  &  4.46  &  2.42     \\
  110220  &  5.60 $^{+0.10}_{-0.10}$    &  1.30$^{+0.00}_{-0.00}$   &  7.28$^{+0.13}_{-0.13}$   &  $944.38\pm0.05$  &  34.77 &  909.61  &  0.76  &  4.77  &  9.39     \\
  110523  &  1.73 $^{+0.17}_{-0.17}$    &  0.60                     &  1.04                     &  $623.30\pm0.06$  &  43.52 &  579.78  &  0.48  &  2.73  &  0.22     \\
  110626  &  1.41 $^{+1.22}_{-0.45}$    &  0.63$^{+1.22}_{-0.12}$   &  0.89$^{+3.98}_{-0.40}$   &  $723.0\pm0.3$    &  47.76 &  675.54  &  0.56  &  3.28  &  0.48     \\
  110703  &  3.90 $^{+2.24}_{-1.85}$    &  0.45$^{+0.28}_{-0.10}$   &  1.76$^{+2.73}_{-1.04}$   &  $1103.6\pm0.7$   &  32.33 &  1071.27 &  0.89  &  5.80  &  3.59     \\
  120127  &  1.21 $^{+0.64}_{-0.25}$    &  0.62$^{+0.35}_{-0.10}$   &  0.75$^{+1.04}_{-0.25}$   &  $553.3\pm0.3$    &  31.82 &  521.48  &  0.43  &  2.39  &  0.20     \\
  121002  &  5.44 $^{+3.50}_{-1.20}$    &  0.43$^{+0.33}_{-0.06}$   &  2.34$^{+4.46}_{-0.77}$   &  $1629.18\pm0.02$ &  74.27 &  1554.91 &  1.30  &  9.28  &  14.94    \\
  130626  &  1.98 $^{+1.20}_{-0.44}$    &  0.74$^{+0.49}_{-0.11}$   &  1.47$^{+2.45}_{-0.50}$   &  $952.4\pm0.1$    &  66.87 &  885.53  &  0.74  &  4.62  &  1.75     \\
  130628  &  0.64 $^{+0.13}_{-0.13}$    &  1.91$^{+0.29}_{-0.23}$   &  1.22$^{+0.47}_{-0.37}$   &  $469.88\pm0.01$  &  52.58 &  417.3   &  0.35  &  1.87  &  0.19     \\
  130729  &  15.61$^{+9.98}_{-6.27}$    &  0.22$^{+0.17}_{-0.05}$   &  3.43$^{+6.55}_{-1.81}$   &  $861\pm2$        &  31    &  830     &  0.69  &  4.24  &  3.35     \\
  131104  &  2.37 $^{+0.89}_{-0.45}$    &  1.16$^{+0.35}_{-0.13}$   &  2.75$^{+2.17}_{-0.76}$   &  $779\pm3$        &  71.1  &  707.9   &  0.59  &  3.50  &  1.72     \\
  140514  &  2.80 $^{+3.50}_{-0.70}$    &  0.47$^{+0.11}_{-0.08}$   &  1.32$^{+2.34}_{-0.50}$   &  $562.7\pm0.6$    &  34.9  &  527.8   &  0.44  &  2.46  &  0.37     \\
  150418  &  0.80 $^{+0.30}_{-0.30}$    &  2.20$^{+0.60}_{-0.30}$   &  1.76$^{+1.32}_{-0.81}$   &  $776.2\pm0.5$    &  188.5 &  587.7   &  0.49  &  2.79  &  0.66     \\
  \hline
  \end{tabular}
  \end{spacing}
    \begin{tablenotes}
       \item[a] $DM$ and $DM_{\rm Galaxy}$ are the total dispersion measure and the contribution from the local Galaxy, repectively.
       \item[b] The dispersion measure excess, which is defined as $DM  - DM_{\rm Galaxy}$.
       \item[c] The redshifts are estimated from the corresponding $DM$ excess. With these redshifts, the luminosity distances ($D_{\rm L}$) and the emitted energies ($E$) can then be calculated. \\
    \end{tablenotes}
  \end{threeparttable}
  }

\end{table*}

\begin{figure*}
\centering
\includegraphics[width=5.8in]{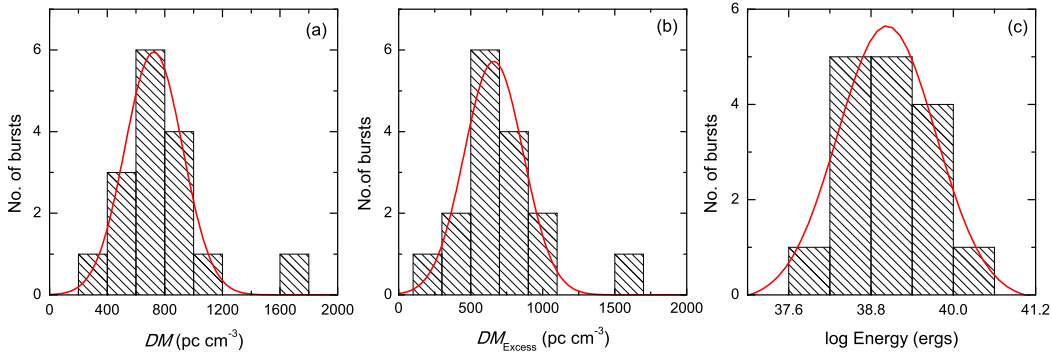}
\caption{Distributions of the DM (Panel a), the DM excess (Panel b),
         and the estimated energy (Panel c). The solid curve in each panel is the best-fit
         Gaussian function, with the fitting correlation coefficients being
         0.90, 0.95 and 0.96 in Panels (a), (b) and (c), respectively. \label{fig1}}
\end{figure*}

\begin{figure*}
\centering
\includegraphics[width=5.8in]{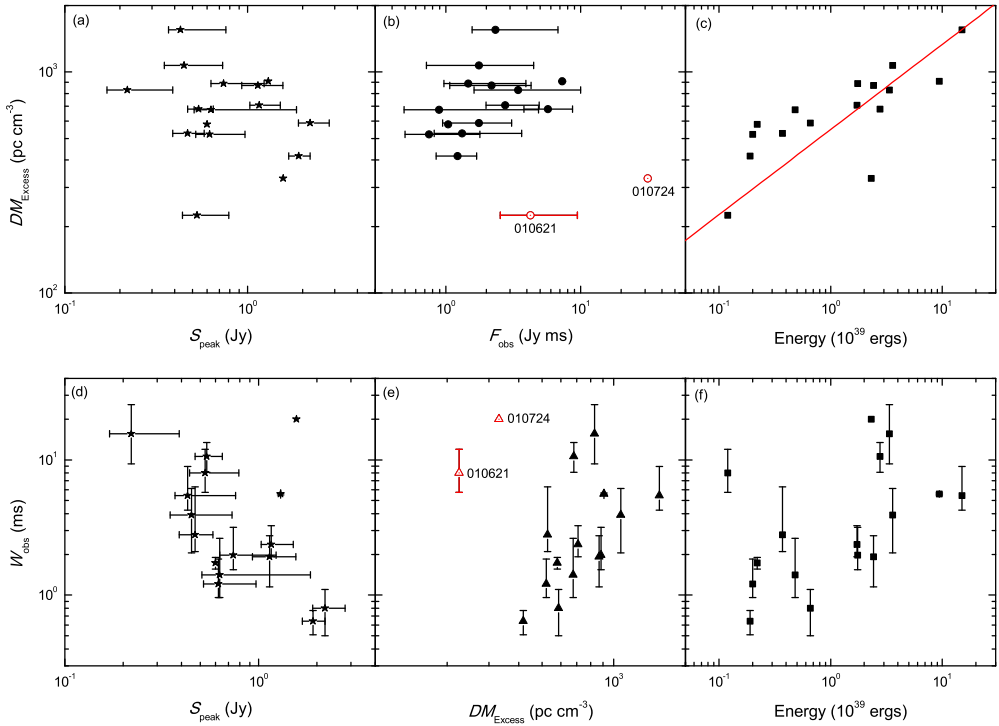}
\caption{Panels (a), (b), and (c) illustrate the observed peak flux density ($S_{\rm peak}$),
         the observed fluence ($F_{\rm obs}$), and the estimated energy ($E$) against the DM excess
         ($DM_{\rm Excess}$), respectively. Panels (d), (e) and (f) present $S_{\rm peak}$, $DM_{\rm Excess}$
         and $E$ against the observed pulse width ($W_{\rm obs}$), respectively.
         The best fit line is shown in Panel (c) when the two parameters are clearly correlated.  \label{fig2}}
\end{figure*}

We first focus on the distribution of the observed DMs of 16 non-repeating FRBs in Table~\ref{tab1}.
Fig.~\ref{fig1} illustrates the histogram of DMs (Panel a) and DM excesses (Panel b).
Both $DM$ and $DM_{\rm Excess}$ roughly follow the normal distribution
and can be well fitted with a Gaussian function. The Gaussian function of DM peaks at
$723\pm45\, \rm pc\,cm^{-3}$, while $DM_{\rm Excess}$ peaks at $660\pm60\, \rm pc\,cm^{-3}$.
The standard deviations of these two Gaussian fitting are comparable and of the magnitude of $140 \, \rm pc\, cm^{-3}$.
We see that $DM_{\rm Excess}/DM \sim 90\%$, which supports FRBs' cosmological origin.
Panel (c) of Fig. \ref{fig1} shows that the estimated radio energy approximatively follows
a log-normal distribution. The log-normal peak is about $10^{39}$ ergs,
consistent with an earlier estimation by \citet{Huang2015}
when only 10 bursts were available.

In Panels (a), (b) and (c) of Fig.~\ref{fig2}, we plot the observed peak flux density, the observed fluence and
the estimated radio energy against the DM excess, respectively.
Fig.~(2a) shows that $S_{\rm peak}$ does not have any clear correlation with $DM_{\rm Excess}$,
which is somewhat unexpected since a more distant source usually tends to be dimmer.
One possible reason is that $S_{\rm peak}$ depends on the time and frequency
resolution of the radio telescope, and another reason may be that the currently observed
$DM_{\rm Excess}$ values are still in a relatively narrow range (the largest $DM_{\rm Excess}$ is
only $\sim 7$ times that of the smallest one, and the estimated $D_{\rm L}$ range is a factor of $\sim$ 10).
Similarly, Fig.~(2b) shows that $F_{\rm obs}$ does not correlate with $DM_{\rm Excess}$.
It indicates that the width of FRBs also does not depend on $DM_{\rm Excess}$,
which will be further shown in the following figures.
In Fig.~(2c), we see that the energies shows a strong correlation with
$DM_{\rm Excess}$, which is natural since the energy emitted has a square dependence on the distance.
In fact, the best fit result of Fig.~(2c) is $E \propto DM_{\rm Excess}^{2.59\pm0.39}$,
with the correlation coefficient and P-value (rejection possibility) as 0.78 and $1.07\times 10^{-5}$, respectively.
It corresponds to a correlation between the energy and the luminosity distance as $E \propto D_{\rm L}^{2.05\pm0.32}$.
The power-law index here is roughly consistent with the square relation within a still relatively large error box.
Note that the correlation may also be partly caused by the telescope selection effect,
because weaker FRB events can be detected only at nearer distances, although they may also happen at far distances.
In addition, the sample of FRBs is still limited.
It is expected that more FRBs would be detected when radio telescopes with higher sensitivities come into operation in the future.

Fig.~(2d) demonstrates the trend for FRBs with brighter peaks ($S_{\rm peak}$) to have narrower width ($W_{\rm obs}$).
A similar tendency has been found for giant pulses from some pulsars
(e.g. Popov \& Stappers 2007; Bhat, Tingay \& Knight 2008; Popov et al. 2009;
Cordes et al. 2016; Popov \& Pshirkov 2016).
In the case of FRBs, this correlation can be explained by some possible models.
For example, \citet{Geng2015} argued that FRBs can be produced by the collisions of asteroids with neutron stars.
In this scenario, when the asteroid collides with the neutron star with a very small impact parameter,
the collision process will finish very quickly and the brightness of the FRB should be high.
On the other hand, if the asteroid collides with the neutron star with a slightly larger impact parameter,
the collision process will be significantly prolonged and the peak flux of the induced FRB
will be correspondingly weaker.
It can naturally account for the correlation of $W_{\rm obs}$ and $S_{\rm peak}$ as shown in Fig.~(2d).
Finally, we see that neither $E$ nor $DM_{\rm Excess}$ correlate with $W_{\rm obs}$ (Figs.~(2e) and (2f)).
Note that while the observed pulse width is relatively clustered, emitted energies span two orders of magnitude.
In Panels (b) and (e), we mark the positions of FRBs 010621 and 010724.
These two events seem to be quite different from others.
We argue that they may form a distinct group, characterized by a low DM and a large fluence.
It suggests the existence of different FRB populations.
More events detected in the future will help to clarify such a possibility.

\section{Intensity Distribution Function}

An absolutely scaled luminosity function can help to reveal the nature of FRBs (Bera et al. 2016).
Since the redshifts of FRBs have not been independently measured,
the derived absolute luminosities and the emitted energies of FRBs are thus controversial
(Katz 2014b; Luan \& Goldreich 2014; Pen \& Connor 2015).
On the other hand, the apparent intensity distribution function of astronomical objects can also
provide helpful information on their nature. A good example is the study of gamma-ray bursts (GRBs).
Before 1997, when the redshifts of GRBs were still unavailable, a deviation from the $-3/2$ power-law
in the peak flux distribution of GRBs was noted (Tavani 1998). It was explained as a hint for
the cosmological origin of GRBs, which was later confirmed by direct redshift measurements.

Here, we focus on the observed fluence ($F_{\rm obs}$) of FRBs, instead of $S_{\rm peak}$.
Seriously affected by scatter and scintillation of IGM,
the peak flux density can be relatively unstable. The combination of $W_{\rm obs}$ and
$S_{\rm peak}$, i.e. the observed fluence, can then more reliably indicate the fierceness of FRBs.
Another reason is that due to the limited time resolving power of our radio receivers, a FRB
should last long enough to be recorded so that the duration is also a key factor.
In fact, a tentative cumulative distribution vs. the fluence has been drawn by
\citet{Katz2016} based on a smaller data set consisting of 10 FRBs.
\citet{Caleb2016} has also tried to compose a cumulative log$N(>F) \,-$ log$F$ correlation by using 9 FRBs
in the high latitude (Hilat) region of the Parkes survey.

\begin{figure*}
\centering
\includegraphics[width=5.2in]{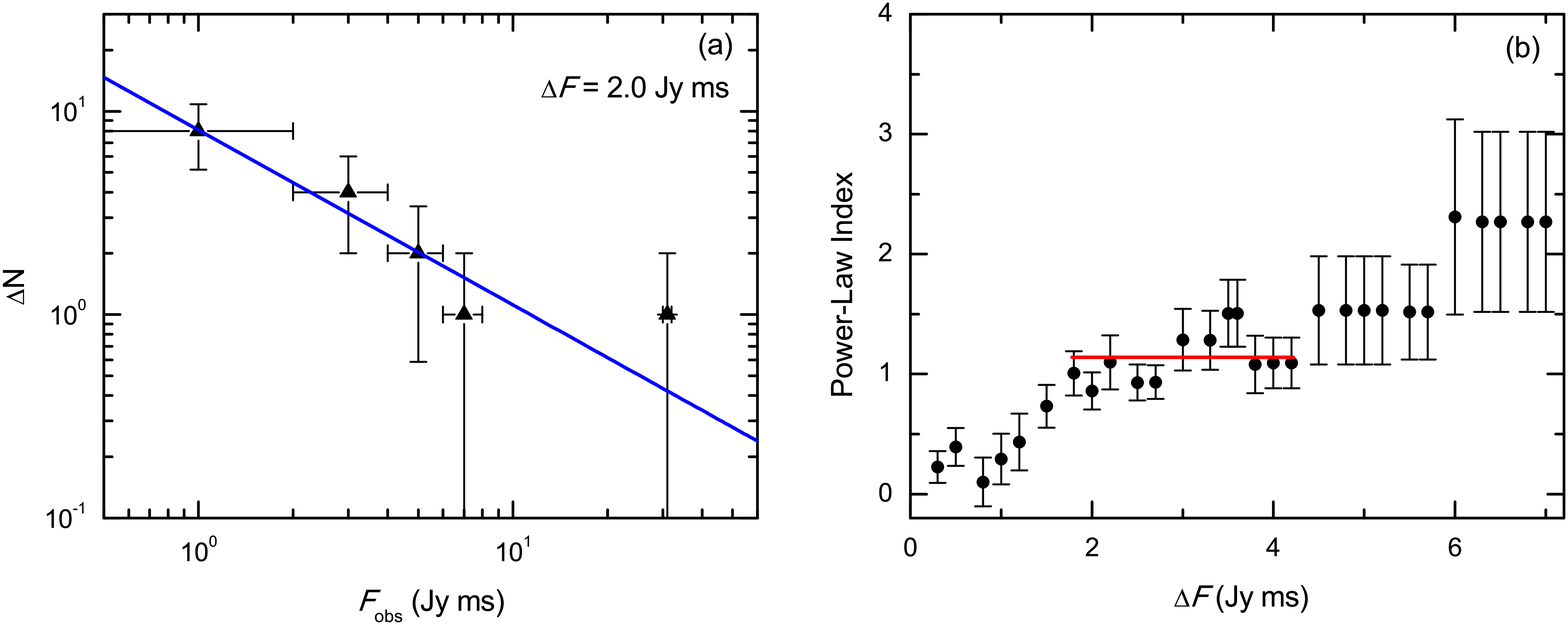}
\caption{Intensity distribution functions of 16 non-repeating FRBs listed in Table~\ref{tab1} (Case I).
         Panel (a) shows an exemplar distribution function
         for a particular bin width, with the $y$-axis showing the number of FRBs in each bin.
         Panel (b) illustrates the best-fitted $a$ values for different bin widths,
         in which the solid short horizontal line shows the average value of $a$
         for a preferable range of $\Delta F$ when $a$ is relatively stable. \label{fig3}}
\end{figure*}

Although the energy distribution of FRBs spans about two orders of magnitude,
it is still relatively clustered,
which indicates that FRBs can be considered as standard candles to some extent.
We can then use the brightness distribution of FRBs to hint their spatial distribution.
We assume that the actual number density of FRBs occurring in the whole sky per day at
a particular fluence $F_{\rm obs}$ follows a power-law function,
i.e. $dN/dF_{\rm obs} = A\, F_{\rm obs}^{-a}$,
where $A$ is a constant coefficient in units of events~$\rm sky^{-1}\,day^{-1}$
and $a$ is the power-law index.
Both $A$ and $a$ need to be determined from observations.
We first only consider the 16 non-repeating FRBs in Table~\ref{tab1} as the input data (Case I).
We group the 16 FRBs into different fluence bins with a bin width of $\Delta F$,
count the number of FRBs in each bin and get the best-fitted power-law
function for $dN/dF_{\rm obs}$. In Panel (a) of Fig.~\ref{fig3}, we show the best-fitted
result when the bin width is taken as $\Delta F = 2.0$ Jy~ms.
The fitted power-law index is $a = 0.86 \pm 0.15$, with the fitted correlation
coefficient being 0.86 and the corresponding P-value being 0.001.
Note that the error bars along $x$-axis simply represent the bin size, and the $y$-axis error
bars are the statistical errors, which are the square root of the number in each bin.

Obviously, since the total number of FRBs is still very limited,
the choice of the bin width will seriously affect the fitting result.
So, we have tried various bin width ranging from 0.3 Jy~ms to 7.0 Jy~ms to study the effect.
For these different bin widths, the derived power-law indices are illustrated
in Panel~(b) of Fig.~\ref{fig3}.
From this panel, we see that when the bin width is very small ($\Delta F \leq 1.8$ Jy~ms),
the fitted $a$ value depends strongly on the bin width. The reason is that only two or three FRBs are
grouped into one bin generally, thus the fluctuation dominates the final result.
Meanwhile, when the bin width is too large ($\Delta F > 4.2$ Jy~ms),
the error bar of the fitted $a$ also becomes larger. In these cases,
only two or three bins are left with a non-zero number of FRBs
so that the derived $a$ becomes unreliable again. On the contrary,
when $\Delta F$ is in the range of 1.8 --- 4.2~Jy~ms, the best-fitted $a$ keeps
to be somehow constant and the error box is also small.
So we choose such a $\Delta F$ range to derive the $a$ parameter.
To reduce the effects of fluctuation as far as possible,
we add up all the 12 $a$ values derived for $\Delta F$ ranging between
$1.8\,{\rm Jy\,ms} \leq \Delta F \leq 4.2 \,{\rm Jy\,ms}$
to get a mean value for $a$ (designated as $a_1$ for Case I), which finally
gives $a_{1} = 1.14 \pm 0.20$ (see Fig.~(3b)).

\begin{table*}
  \centering
  \caption{The observable event rate of FRBs in the literature.  \label{tab2}}
  \begin{tabular}{@{}cccc@{}}
  \hline
  $F_{\rm Limit}$    &  $R(>F_{\rm Limit})$            & Reference &  Derived coefficient $A$   \\
   (Jy~ms)   &  ($\rm sky^{-1}\,day^{-1}$) &           & $(10^{3}\, \rm sky^{-1}\,day^{-1}$) \\
  \hline
      3.0  &    $10^{4}$                  & \citet{Thornton2013}   & $5.61\pm2.04$    \\
      0.35 &    $3.1\times 10^{4}$        & \citet{Spitler2014}    & $7.88\pm6.92$    \\
      2.0  &    $2.5 \times 10^{3}$       & \citet{Keane2015}      & $1.17\pm0.51$    \\
      1.8  &    $1.2\times 10^{4}$        & \citet{Law2015}        & $5.37\pm2.47$    \\
      4.0  &    $4.4\times 10^{3}$        & \citet{Rane2016}       & $2.86\pm0.90$    \\
  0.13-5.9 &    $6.0 \times 10^{3}$       & \citet{Champion2015}   & $1.94\pm1.27$    \\
  \hline
  \end{tabular}
\end{table*}

\begin{figure*}
\centering
\includegraphics[width=5.2in]{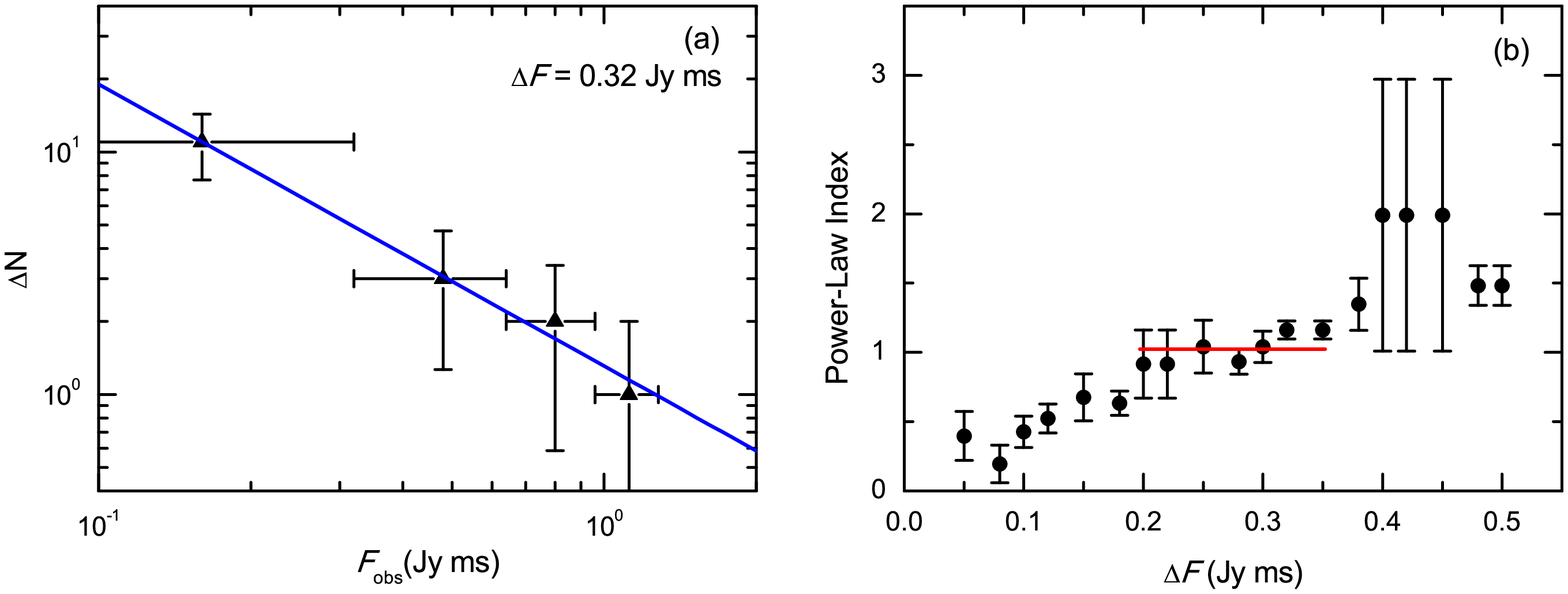}
\caption{Intensity distribution functions of the 17 repeating FRBs from the source of FRB 121102 (Case II).
         Symbols are the same as Fig. 3.
         \label{fig4}}
\end{figure*}

Integrating the intensity distribution function, we can derive the FRB event rate
above a particular fluence limit in the whole sky per day as
\begin{equation}
  R(> F_{\rm Limit})=A\int^{F_{\rm max}}_{F_{\rm Limit}}F_{\rm obs}^{-a}\,dF_{\rm obs},
\end{equation}
where $F_{\rm Limit}$ corresponds to the limiting sensitivity of a radio telescope,
$F_{\rm max}$ is the upper limit of the fluence of observed FRBs, which is set as 35 Jy~ms
in our calculations (the observed maximum fluence is $\sim 32$ Jy ms at present).
$A$ is an unknown coefficient. It still needs to be determined from observations.
FRBs were mainly identified from the archival data of several radio surveys. Constraints
on the actual event rates of FRBs above a certain fluence limit were also derived in
these analyses and were reported in the literature. We sum up these constraints in
Table~\ref{tab2}. According to Eq.~(1), these data can be used to derive the coefficient $A$ by
using our best-fit $a_{1}$ value. The results are also listed in the last column in Table~\ref{tab2}.
Combining all these different $A$ values, we finally get a mean value as
$A = (4.14 \pm 1.30) \times 10^3 \, \rm sky^{-1}\,day^{-1}$.
After getting the power-law index $a$ and the constant coefficient $A$, we now can write down
the full apparent intensity distribution function as,
\begin{equation}
\frac{dN}{dF_{\rm obs}} = (4.14 \pm 1.30) \times 10^3\, F_{\rm obs}^{-1.14 \pm 0.20} \, \rm sky^{-1}\,day^{-1},
\end{equation}
where $F_{\rm obs}$ is in units of Jy~ms.

In total, 17 repeating FRB events have been detected from the source of FRB 121102.
We treat these events as a separate group (Case II) and also study their intensity distribution.
Similar fitting processes as for Case I are applied to Case II, the final results are shown in Fig.~\ref{fig4}.
In Fig.~(4a), we show the exemplar fitting result when taking the bin
width as $\Delta F=$ 0.32~Jy~ms. The best-fitted power-law index is $1.16\pm0.06$, with the correlation
coefficient and P-value being 0.99 and 0.001, respectively. Similar to Fig.~(3b), Fig.~(4b) shows the
derived $a$ values for different bin widths. The most probable mean value of $a$ (designated as $a_2$ for
Case II) is calculated for $\Delta F$ ranging between 0.2~Jy~ms and 0.35~Jy~ms, which is $a_{2}=1.03\pm0.16$.
Note that although $a_1$ and $a_2$ do not differ from each other markedly, there are
actually significant differences between the overall characteristics of non-repeating FRBs and repeating FRBs.
For example, the most obvious difference is that the repeating FRBs are generally weaker, indicating
that they are mainly at the weak end of the fluence distribution.

\section{Prospects for FAST}

FAST (Nan et al. 2011; Li, Nan \& Pan 2013), an ambitious Chinese mega-science project, is currently being
built in Guizhou province in southwestern China. With a geometrical diameter of
500 meters and an effective diameter of $\sim 300$ meters at any particular moment,
it will be the largest single-dish radio telescope in the world when it comes into
operation in September of 2016. FAST's receivers will cover both low frequency (70-500 MHz)
and middle frequency (0.5-3 GHz) bands.
We here consider the prospect of detecting FRBs with FAST by using
the derived intensity distribution function.
Our calculations are done at the L band (1400 MHz) of FAST, which
is the central observational frequency for most detected FRBs.

The sensitivity or the limiting flux density ($S_{\rm limit}$) of a radio telescope can
be estimated as (Zhang et al. 2015),
\begin{equation}
\begin{aligned}
S_{\rm limit}\simeq \, & (12\mu {\rm Jy})(\frac{0.77\times10^3 {\rm m^2/K}}{A_{\rm e}/T_{\rm sys}})
                          (\frac{S/N}{3})(\frac{\rm 1 hour}{\Delta \tau})^{1/2} \\
                       &(\frac{\rm 100 MHz}{\Delta \nu})^{1/2}, \\
\end{aligned}
\end{equation}
where $T_{\rm sys}$ is the system temperature, $\Delta \tau$ is the integration time,
$\Delta \nu$ is the bandwidth, $\rm S/N$ is the signal-to-noise ratio which is usually
taken as 10 for a credible detection of a FRB (Champion et al. 2016), $A_{\rm e}$ is the
effective area and it equals to $\eta_{A}\,\pi\,(d/2)^{2}$, with $\eta_{A}$
being the aperture efficiency and $d$ being the illuminated diameter. FAST has a system
temperature of $T_{\rm sys} \sim 25 K$ at 1400 MHz. For other parameters of FAST,
we take $d=300\, \rm m$, $\eta_{A}=0.65$ (Zhang et al. 2015), $\Delta \nu = 300$ MHz,
$\Delta \tau = 3 \; {\rm ms}$ (Law et al. 2015). The limiting fluence of FAST is then
calculated to be $F_{\rm Limit} = S_{\rm limit} \times \Delta \tau = 0.03 $ Jy~ms.
Note that from Equations (1) and (2), the actual FRB event rate above the fluence limit
of 0.03 Jy~ms is $(3.03\pm1.56) \times 10^{4}\, \rm sky^{-1}\,day^{-1}$.

The beam size of a radio telescope is $\Omega \sim \pi\theta^{2}$,
where $\theta \sim 1.22(\lambda/d)$ is the half opening angle of the beam. For FAST,
the beam size is $\Omega \sim 0.008 \; {\rm deg}^2$ at 1400 MHz. FAST's reciver has 19 beams
in L band, and the corresponding total instantaneous field-of-view (FoV) is $0.15 \; {\rm deg}^2$.
Considering Eq.~(2) and all these parameters, we can get the daily detection rate of FRBs by
FAST as,
\begin{equation}
\begin{aligned}
R_{\rm FAST} & \sim \rm (3.03\pm1.56) \times 10^{4} \times \frac{0.15 \; deg^{2}}{41253 \; deg^{2}} \; day^{-1}  \\
             & = 0.11 \pm 0.06 \, \rm day^{-1}. \\
\end{aligned}
\end{equation}
For a 1000 hours of observation time, this corresponds to $\sim 5 \pm 2$ detections.

\section{Discussion and Conclusions}

In this study, we analysed statistically the key parameters of the 16 non-repeating FRBs detected by Parkes and GBT.
The observed DM spans a range of 375 --- 1629 ${\rm \,pc\,cm^{-3}}$ and
peaks at $\sim 723 {\rm \,pc\,cm^{-3}}$, while the DM excess peaks at $\sim 660 {\rm \,pc\,cm^{-3}}$
and typically accounts for $\sim 90\%$ of the total DM. The emitted radio energy spans about
two orders of magnitude, with the mean energy being about $10^{39}$ ergs.
While most of the parameters do not correlate with each other,
a burst with stronger $S_{\rm peak}$ tends to have shorter duration.
Meanwhile, a clear correlation between the radio energy released and the DM excess
has been found to be $E \propto DM_{\rm Excess}^{2.59\pm0.39}$ (Fig.~2c),
which may reflect the square dependence of the emitted energy on the distance.
But note that the observational selection effect may also play a role in the statistics.
From these statistical analyses, we found that FRBs 010621 and 010724 are
quite different from others and they may form a distinct group.

Using the observed fluence as an indication for the strength of FRBs and
combining the constraints on the event rate of FRBs from various observational surveys,
we derived an apparent intensity distribution function for the 16 non-repeating FRBs as
$dN/dF_{\rm obs} = (4.14\pm1.30) \times 10^3\, F_{\rm obs}^{-1.14 \pm 0.20} \, \rm sky^{-1}\,day^{-1}$.
For FRB 121102 and its repeating bursts, the corresponding power-law index is derived to be $a_{2}=1.03\pm0.16$.
Based on the intensity distribution function, we were able to estimate the detection rate
of FRBs by China's coming FAST telescope. With a sensitivity of 0.03 Jy~ms and a total
instantaneous FoV of 0.15 deg$^2$ (19 beams), FAST will be able to detect roughly 1 FRB
for every 10 days of observations, or about 5 events for every 1000 hours.

A few authors have studied the cumulative distribution function of FRBs
(Bera et al. 2016; Caleb et al. 2016; Katz 2016; Wang \& Yu 2016), which is
usually assumed to be a power-law function of $N_{>F_{\rm obs}} \propto F_{\rm obs}^{-\alpha}$.
For standard candles distributed homogeneously in a flat Euclidean space, the value of $\alpha$ should
be 3/2 (Thornton et al. 2013).
\citet{Opper2016} has argued that the range of $\alpha$ may be $0.9\leq \alpha \leq 1.8$,
\citet{Caleb2016} has derived $\alpha=0.9$ for a small sample of 9 Hilat FRBs,
while \citet{Wang2016a} considered $\alpha$ as $0.78$ for FRB 121102 and its 16 repeating bursts.
Note that the relation between the cumulative index $\alpha$ and our intensity distribution index
$a$ is $\alpha = a-1$. So, our derived index of $a=1.14\pm 0.20$ will correspond to $\alpha = 0.14 \pm 0.20 $.
It is significantly smaller than Caleb et al.'s value. The difference could be caused by different
sample capacity. We have 16 non-repeating FRBs in our sample.
It strongly points toward a deficiency of the dimmest FRBs, which has also been indicated in
earlier studies (Bera et al. 2016; Caleb et al. 2016; Lyutikov, Burzawa \& Popov 2016).
There are a few factors that could lead to such a deviation. First, the total number of currently
observed FRBs is still very limited. It can result in a large fluctuation in the measured power-law index.
In fact, \citet{Caleb2016} have estimated that at lease $\sim 50$ FRBs are needed to extract conclusive
information on the physical nature of FRBs. Second, FRBs are not ideal standard candles.
But as long as the characteristics of FRBs
does not evolve systematically with the distance, the index will not be affected too much. A wider
brightness range will only result in a larger error box for $\alpha$, which can still be reduced by
increasing the FRB samples. Third, FRBs may not be homogeneous sources, the co-moving density
or their brightness may evolve in space. For example, the co-moving FRB density may be smaller when
the distance increases, or farther FRBs may not be as fierce as those nearer to us.
Fourth, the space may not be a flat Euclidean space, such as for a curved $\Lambda$-CDM
cosmology. In this case, the deviation of $\alpha$ from 3/2 can be used as a probe for the
cosmology (Caleb et al. 2016). Finally, it should also be noted that the apparent deficiency
of the dimmest FRBs could also be due to the selection effect. Much weaker FRB events
may actually have happened in the sky, but we were not able to record them or find them due to
current technical limitations. To make clear which of the above factors has led to the
smaller $\alpha$, more new FRB samples would be necessary in the future.

As addressed above, the apparent intensity distribution function derived here is still a preliminary
result. We need much more samples to determine the power-law index more accurately.
With an enormous effective area for collecting radio emissions, the Chinese FAST telescope
is very suitable for FRB observations. It may be able to increase the FRB samples at a
rate of $\sim 10$ events per year (assuming an effective observation time of 2000 hours).
More importantly, FAST can operate in a very wide frequency range and can hopefully provide
detailed spectrum information for FRBs. It is expected to be a powerful tool in the field.

At present, whether FRBs are beamed or not is still an unclear but important problem.
If FRBs are highly collimated, the actual energy released will be much smaller than
the currently estimated energy base on an assumed solid angle of $\sim 1\,\rm sr$ (Huang \& Geng 2016).
The emission mechanisms of FRBs will then involve complex jet effects (Borra 2013).
Studying the jet effects of FRBs will be an important task and it will help us understand
the explosion mechanisms of FRBs.
When more FRBs are observed and localized, we may be able to get useful information on the beaming effects from direct observations.

\normalem
\begin{acknowledgements}

This study was supported by the China Ministry of Science and Technology under State Key
Development Program for Basic Research (973 program, Grant Nos. 2014CB845800, 2012CB821802),
the National Natural Science Foundation of China (Grant Nos. 11473012, U1431126, 11263002)
and the Strategic Priority Research Program (Grant No. XDB09010302).

\end{acknowledgements}

\label{lastpage}
\end{document}